\documentstyle[12pt]{article}
\newcommand{\reals}{\mbox{${\rm I\!R }$}}
\newcommand{\nats}{\mbox{${\rm I\!N }$}}

\newcommand{\ant}{\int\limits}
\newcommand{\sneu}{\sum_{n=1}^{\infty}}
\newcommand{\sleu}{\sum_{l=1}^{\infty}}

\newcommand{\snnu}{\sum_{n=0}^{\infty}}

\newcommand{\cam}{{\cal M}}

\newcommand{\beq}{\begin{eqnarray}}
\newcommand{\eeq}{\end{eqnarray}}
\newcommand{\nn}{\nonumber}

\newcommand{\abl}{\partial}

\newcommand{\zh}{\zeta_H(s;a)}
\newcommand{\cont}{\ant_{c-i\infty}^{c+i\infty}d\alpha\,\,\Gamma(\alpha)}
\newcommand{\zpi}{2\pi i}
\newcommand{\tint}{\ant_0^{\infty}dt\,\,t^{s-1}}
\newcommand{\epa}{E^{\beta}(s)}

\begin{document}

\def\bce{\begin{center}}
\def\ece{\end{center}}
\def\bea{\begin{eqnarray}}
\def\eea{\end{eqnarray}}
\def\brr{\begin{array}}
\def\err{\end{array}}
\def\ben{\begin{enumerate}}
\def\een{\end{enumerate}}
\def\bei{\begin{itemize}}
\def\eei{\end{itemize}}
\def\ul{\underline}
\def\ni{\noindent}
\def\bs{\bigskip}
\def\ms{\medskip}
\def\dsp{\displaystyle}
\def\wt{\widetilde}
\def\wh{\widehat}
\def\rl{$\Real$}
\def\tu{\bigtriangleup}
\def\td{\bigtriangledown}

\hfill UB-ECM-PF 94/8

\hfill March 1994

\vspace*{3mm}

\begin{center}

{\LARGE \bf
Applications of the Mellin-Barnes integral representation}

\vspace{4mm}

\renewcommand
\baselinestretch{0.8}
\ms

\renewcommand
\baselinestretch{1.4}
{\sc E.~Elizalde}\footnote{E-mail:
eli@zeta.ecm.ub.es} \\
Center for Advanced Studies, C.S.I.C., Cam\'{\i} de Santa
B\`arbara,\\ 17300 Blanes, \\
and
Department E.C.M. and I.F.A.E., Faculty of Physics,
University of \\Barcelona, Diagonal 647, 08028 Barcelona,
Catalonia, Spain,\\
{\sc K.~Kirsten}\footnote{Alexander von
Humboldt Foundation Fellow. E-mail: klaus@ebubecm1.bitnet}
\\  Department E.C.M., Faculty of Physics,
University of  Barcelona, \\  Diagonal 647, 08028 Barcelona,
Catalonia,
Spain \\
{\sc S.~Zerbini}\footnote{E-mail: zerbini@itncisca.bitnet}\\
Dipartimento di Fisica, Universit\`a di Trento, Italia\\
Istituto Nazionale di Fisica Nucleare, Gruppo Collegato di Trento\\

\renewcommand
\baselinestretch{1.4}

\vspace{5mm}

{\bf Abstract}

\end{center}

We apply the Mellin-Barnes integral representation to several
situations of
interest in mathematical-physics. At the purely mathematical level, we
derive useful asymptotic
expansions of different zeta-functions and partition functions. These
results are then employed in different topics of quantum field theory,
which include the high-temperature expansion of the free energy of a
scalar field in ultrastatic curved spacetime, the asymptotics of the
$p$-brane density of states, and an explicit approach to the asymptotics
of the determinants that appear in string theory.

\vspace{4mm}


\newpage

\section{Introduction}

Expansions in terms of asymptotic series constitute a very important
tool in various
branches of physics and mathematics. For example asymptotic expansions
in inverse powers of large masses of the field have always been a
usually employed
approximation (see, for example, \cite{ball,1} and references therein).
Another example is the high-temperature expansion in different contexts
\cite{haberweldon82,imagine}.
Also for the numerical analysis of several functions
the use of asymptotic expansions has been shown to be very powerful
\cite{2,3}.
Finally, let us mention applications to the theory of partitions
\cite{sergio}, where
asymptotic series are basic in the proof of the theorem of Meinardus
\cite{meinardus54,mein}, which
is fundamental in the calculation of the asymptotic  density of
$p$-brane states \cite{bytsenkokirstenzerbini93}.

In all the mentioned applications, the asymptotic series arises when
doing a calculation of a functional determinant, which has to
be regularized by some method.
Usually, this functional determinant is a one-loop approximation to a
functional integral resulting from integrating out the quadratic part
of the quantum fluctuations around some background fields extremizing
the
action \cite{sergei}. The quadratic part of the quantum fluctuations is
often described by elliptic operators of the form
\beq
E=-D^2 +M^2,\qquad D_{\mu} =\partial _{\mu} -iA_{\mu},\label{elliptic}
\eeq
with a gauge potential $A_{\mu}$ and some (effective) mass $M^2$ of the
theory.

One
of the possible regularization schemes for the product of eigenvalues is
the zeta-function regularization procedure introduced some
time ago \cite{hawking77}. In this procedure the one-loop quantum
correction $\Gamma$ to the action is described by
\beq
\Gamma =\frac 1 2 \ln\det E =-\frac 1 2 \left[\zeta'_E(0)
-\ln\mu^2\zeta_E (0)\right],\nn
\eeq
being $\zeta_E (s)$ the zeta-function associated with the operator $E$,
that is
\beq
\zeta_E (s) =\sum_j\lambda_j^{-s},\nn
\eeq
where $\lambda_j$ are the eigenvalues of $E$. Furthermore, $\mu$ is a
mass
scale that needs to be introduced for dimensional reasons. Thus, in
using this scheme, the basic mathematical tool is the analysis of
zeta-functions associated with (pseudo-) elliptic differential operators
and the determination of its properties depending on the problem under
consideration.

On the purely mathematical side of the problem a very interesting and
self-contained reference is the one by Jorgenson and Lang \cite{lang}.
In these lecture notes the authors describe how parts of analytic number
theory and parts
of the spectral theory of certain operators can be merged under a
more general analytic theory of regularized products of certain
sequences of numbers satisfying a few basic axioms. However, their
exposition is kept on a rather general level, which is not directly
applicable
to the physical problems. Physicists have often to deal with very
specific situations, generally fulfilling these few basic axioms, but
for such specific
situations it is most useful if results are given in full
detail. There thus
appears the need to tend a bridge across the purely mathematical
part of the problem and its specific applications in physics. This is
part of the motivation of the present article, its main emphasis being
on the physical applications described below.

In order to keep it self-contained, we have decided to refer not only to
the
associated mathematical literature, but also to present the techniques
in the context of physically relevant situations.
So, for example, a detailed treatment of Epstein-type zeta-functions is
of great interest, because these functions are essential for the
computation of effective actions in non-trivial backgrounds that
appear in different contexts \cite{topmass} and for the analysis of the
Casimir effect \cite{casimir}.

The most elementary example is that of the Hurwitz zeta-function
\beq
\zeta_H(s;a)=\snnu (n+a)^{-s}\nn
\eeq
which has been treated in detail only recently in
\cite{elizalde86,rudaz90,elizalde93}. Before, the
only derivatives of $\zeta_H(s;a)$ available in the usual tables were
\beq
\frac{\abl}{\abl a}\zeta_H(s;a)&=&-s\, \zeta_H(s+1;a),\nn\\
\left. \frac{\abl}{\abl s}\zeta_H(s;a)\right|_{s=0}&=&\ln\Gamma(a)-\frac
1 2 \ln(2\pi). \nn
\eeq
The aim in \cite{elizalde86} was to obtain an asymptotic expansion of
$\abl \zeta_H(s;a)/\abl s$ valid for all negative values of $s$ using as
starting point Hermite's integral representation for $\zeta_H(s;a)$.
Later on, this result has been rederived using a direct method connected
with the calculus of finite differences \cite{rudaz90}, applicable also
to other functions of interest. The result has been extended to $\abl
^n\zeta_H(s;a)/\abl s^n$ in \cite{elizalde93}, where a very useful
recurrent formula has been derived, which completely solves the problem
of the calculation of any derivative of the Hurwitz zeta-function.
In all these references, it has been
mentioned that the asymptotic expansion of $\zeta_H'(s;a)$ may be also
simply found by differentiation of the asymptotic expansion of
$\zeta_H(s;a)$, what is, however, a procedure not justified a priori
(in fact, it is controlled by a {\it tauberian} theorem).

In order to introduce and illustrate a different, powerful
technique for the
derivation of asymptotic series, which is the Mellin transformation
technique (Jorgenson and Lang call it also vertical transform), we
will take once
more the Hurwitz zeta-function $\zeta_H(s;a)$. It serves as an
elementary example in order to introduce the Mellin technique, which,
as mentioned, is
applicable in a much wider range \cite{lang}. We will
derive
an asymptotic series expansion for $\abl^n
\zeta_H(s;a)$ $/\abl s^n (s;a)$, for all values $n\in\nats_0$.
Especially we will
show again that, in order to obtain the asymptotic series of the
derivatives
of the Hurwitz zeta function, one may simply differentiate the
asymptotic expansion (a method that was proven to be valid in this case
by applying standard mathematical results for asymptotic
expansion,
such as Laplace's method and Watson's lemma \cite{elizalde86,2}).

After having presented
the
method for the case of $\zeta_H(s;a)$, we consider other different
quantities, explaining briefly at the beginning of each section how
these appear in actual physical situations.
In
Sect.~3 the more complicated example of an Epstein-type zeta-function
is presented. Sect.~4 is devoted to the treatment of sums which most
frequently
appear in finite-temperature quantum field theory and in the theory of
partitions. We use the new technique in order to rederive the
high-temperature expansion of a free scalar field in curved spacetime.
Furthermore, we outline a generalization of the theorem of
Meinardus
\cite{meinardus54,mein}, which enables one to
find the asymptotic state density of
$p$-branes \cite{bytsenkokirstenzerbini93}. The
last application is concerned with properties of
some determinants appearing in (super-) string theory.
In the conclusions of the paper, the results
presented here are briefly summarized.
\section{Asymptotic expansion of the Hurwitz zeta-function}
The aim of this section is to explain how the approach works in
obtaining the asymptotic expansions of a large class of functions
\cite{lang}. As an
example, we choose the very useful case of the Hurwitz zeta-function
$\zh$, defined by \cite{erdelyimagnusoberhettingertricomi53}
\beq
\zh =\snnu (n+a)^{-s}\label{h1}, \ \ \ 0 < a \leq 1,  \ \ \ \Re \, s >1,
\eeq
and derive an asymptotic expansion for large values of $a$
(for previous treatments see \cite{elizalde86,rudaz90,elizalde93}).

To start with, we rewrite equation (\ref{h1}), as usually, in the
form
\beq
\zh =a^{-s}+\frac 1 {\Gamma(s)}\sneu\tint e^{-(n+a)t}.  \nn
\eeq
The key idea is to make use of the complex integral representation
of the exponential in the form of an integral of Mellin-Barnes type,
\beq
e^{-v}=\frac 1 {\zpi}\cont v^{-\alpha}, \label{h2}
\eeq
with $\Re v>0$ and $c\in\reals$, $c>0$. Restricting it to the
part $e^{-nt}$ only, it leads to
\beq
\zh&=&a^{-s}+\frac 1 {\zpi \Gamma(s)}\sneu \tint e^{-at}\cont
n^{-\alpha}t^{-\alpha}\nn\\
&=&a^{-s}+\frac 1 {\zpi \Gamma(s)}\sneu \,\,\,\cont\Gamma(s-\alpha )
   n^{-\alpha}a^{\alpha-s}. \label{h3}
\eeq
Now we would like to interchange the summation and integration in order
to arrive at an expression in terms of Riemann zeta-functions.
By choosing $c>1$, the resulting sum is absolutely convergent, leading
to \beq
\zh=a^{-s}+\frac 1 {\zpi \Gamma(s)}\cont\Gamma(s-\alpha )
   \zeta_R(\alpha)a^{\alpha-s}.  \label{h4}
\eeq
The integrand has poles on the left of the contour at (choosing $s>c$)
$\alpha =1$ (the pole of $\zeta_R (\alpha))$ and $\alpha=-k$,
$k\in\nats_0$
(the poles of $\Gamma (\alpha)$). All poles are of order one and, with
$\zeta
(1-2m)=-B_{2m}/(2m)$, for $m\in\nats$, one easily finds the asymptotic
behaviour \beq
\zh\sim \frac 1 2 a^{-s}+\frac 1 {s-1}a^{-s+1}
+\sum_{k=2}^{\infty}\frac{(s)_{k-1}B_k}{k!}a^{-s-k+1}, \label{h5}
\eeq
where $(s)_k$ is the Pochhammer symbol $(s)_k =\Gamma(s+k)/\Gamma(s)$.
This result
agrees (of course) with the known result. However, here it has been
rederived with nearly no calculational effort.

Let us now concentrate on the asymptotic expansion of $\frac{\abl
^n}{\abl s^n}\zh$. It has already been proven in the literature
\cite{elizalde86,rudaz90,elizalde93}, that the asymptotic expansion of
the first derivative of the Hurwitz zeta-function is simply the term by
term derivative of the asymptotic expansion (\ref{h5}), a procedure
which is not justified a priori and needs a lengthy demonstration in
terms of Watson's lemma. We would now like to show, that in the
case considered above our procedure is true for {\it all} derivatives of
$\zh$.

The proof is the following. Looking at (\ref{h4}), it is easily
seen that integration
and differentiation may be savely commuted. The reason is, that
differentiation does not destroy the rapid decay of the gamma
function, which is seen using the representation 8.341.1 in
\cite{gradshteynryzhik65} for $\ln\Gamma (z)$. In fact, by doing so no
additional
poles are created and the residue of the pole is just the derivative of
the old one. Thus the asymptotic expansion of $\frac{\abl ^n}{\abl
s^n}\zh$ is simply the term by term differentiate of the equation
(\ref{h5}). Formulas for very quick explicit derivation of those
(together with some basic examples) can be found in \cite{elizalde93}.
\section{Asymptotic series expansion of Epstein-type zeta-functions}
\setcounter{equation}{0}
As the next example, we would like to consider the Epstein-type
zeta-function
\beq
E^2(s)=\snnu [(n+a)^2+M^2]^{-s}.\label{basic}
\eeq
The range of summation is chosen in a way that $M=0$ corresponds to the
Hurwitz zeta-function. As the presented calculations will show, other
index ranges (depending on the details of the subsequently described
physical situations) do not give rise to additional problems and may be
treated in exactly the same way.

This type of function, or multidimensional generalizations of it, is of
importance for different problems of quantum
field theory. For example it naturally appears in the context of gauge
field mass generation in partially compactified spacetimes of the type
$T^N \times \reals^n$. In some detail, a massive complex scalar field
$\phi$ defined on $T^N \times \reals ^n$ with, for example, periodic
boundary conditions for each of the toroidal components is coupled to a
constant Abelian gauge potential $A_{\mu}$. Due to the non-trivial
topology, constant
values of the toroidal components are physical parameters of the theory
and the effective potential of the gauge theory will depend on these
parameters. For the calculation of the effective potential a detailed
knowledge of $E^2 (s)$ is necessary, where the parameter $a$ corresponds
to an Abelian gauge potential in a toroidal dimension
\cite{actor,david}. The parameter $M^2$ here plays the role of the mass
squared $m^2$ of the scalar field. Realizing that an imaginary constant
gauge potential is equivalent to a
chemical potential \cite{imagine}, the relevance of $E^2 (s)$
for
finite temperature quantum field theory in Minkowski spacetime and for
the phenomenon of Bose-Einstein condensation is also obvious
\cite{haberweldon82}.
The case $a=0$ is relevant to describe topics like topological symmetry
breaking or restoration in self-interacting $\lambda \phi ^4$ scalar
field theories on the spacetime $\reals ^3\times S^1$ \cite{topmass}.
There, the effective mass naturally appearing in the theory is $M^2 =m^2
+(\lambda/2)\hat{\phi}^2$, with the classical scalar background field
$\hat{\phi}$. Finally,
let us mention the appearance of similar functions in Casimir energy
calculations in quantum field theory in spacetimes with compactified
dimensions \cite{casimir}.

In all the above mentioned problems, the way the dimensions are
compactified (circle, parallel plates) and the relevant boundary
conditions for the field (periodic, antiperiodic, Neumann, Dirichlet),
may lead to a different range of summations in (\ref{basic}).

As already mentioned, the parameter
$M^2$ usually plays the role of an (effective) mass of the theory.
An approximation often employed is
the
large mass approximation in which one looks for an asymptotic expansion
of physical quantities in inverse powers of the mass. We will now show
that using the approach of section 2 this may be very easily obtained
too. For the sake of generality let us consider
\beq
\epa=\snnu [(n+a)^{\beta}+M^2]^{-s},\label{h6}
\eeq
where the arbitrary (positive) $\beta$ leads to no additional
complications and is included for that reason.

We are interested in the asymptotic expansion for large values of $M^2$.
Proceeding as in section 2, using the Mellin transform of the
exponential, this time for $e^{-(n+a)^{\beta}}$, one arrives at
\beq
\epa&=&\frac 1 {\zpi \Gamma(s)}\snnu\tint e^{-M^2t}\cont
(n+a)^{-\beta\alpha}t^{-\alpha}\nn\\
&=&\frac 1 {\zpi \Gamma(s)}\cont \Gamma(s-\alpha)\zeta_H(\beta\alpha;a)
M^{2\alpha-2s}.  \label{h7}
\eeq
Once more, the only poles that appear are of order one, located
at $\alpha =\frac 1 {\beta}$ (pole of the Hurwitz zeta function) and at
$\alpha=-n$, $n\in\nats_0$ (poles of the Gamma function).
The residues are easily found, leading to the asymptotic series
\beq
\epa&\sim &\frac{\Gamma\left(s-\frac 1
{\beta}\right)}{\beta\Gamma(s)}\Gamma\left(\frac 1
{\beta}\right)M^{2\left(\frac 1 {\beta}-s\right)}\nn\\
& &+\frac 1 {\Gamma(s)}\snnu (-1)^n\frac{\Gamma(s+n)}{n!}\zeta_H(-\beta
n;a)M^{-2s-2n}.\label{h8}
\eeq
For the special case $\beta =m\in\nats$, using \cite{gradshteynryzhik65}
\beq
\zeta_H(-n;a)=-\frac{B_{n+1}(a)}{n+1}, \nn
\eeq
with the Bernoulli polynomials $B_n(a)$,
Eq. (\ref{h8}) may be written as
\beq
E^m(s)&\sim &\frac{\Gamma\left(s-\frac 1
{m}\right)}{m\Gamma(s)}\Gamma\left(\frac 1
{m}\right)M^{2\left(\frac 1 {m}-s\right)}\nn\\
& &-\frac 1 {\Gamma(s)}\snnu (-1)^n\frac{\Gamma(s+n)}{n!}
\frac{B_{m n+1} (a)}{m n+1} M^{-2s-2n}.
\label{h9}
\eeq
Expressions of this sort (at least in the most common case when $\beta
=2$) have been applied in the past in models of effective
Lagrangians with the aim at understanding the quark confinement problem
in QCD \cite{1}. In those theories, the heavy mass limit for the quarks
appears naturally as a good
(in principle) first approximation. Different models along the same line
are now fashionable again and expressions of the kind (\ref{h9}) may
prove to be useful once more. On the other hand, Eq. (\ref{h9}) and, in
particular, its partial derivatives with respect to $s$ and $M^2$ are
the sort of basic expressions that appear in other physical
applications where compactification of spacetime plays a basic role,
apart from the one already mentioned, these are for example
Kaluza-Klein theories \cite{4} and the computation of the vacuum energy
density in compact or cylindrical universes \cite{5}.

It is rather obvious how to change the procedure in order to obtain
a corresponding asymptotic expansion for the generalization  of
(\ref{h6}) in the form of a
multidimensional series (see \cite{6}, where this idea is developed in
detail). Whenever
this approach is useful, the calculational effort (if we only
keep,
as here, the dominant terms of the asymptotic expansion) will not
exceed the one presented in the two examples above.
The corresponding results are of relevance, whenever in the context of
topological mass generation \cite{topmass} or the Casimir effect
\cite{casimir} more than one dimension is compactified.

\section{Application to partition sums}
\setcounter{equation}{0}
In order to get a more detailed idea about how to use the presented
techniques, let us consider the quantity
\beq
G(t)=\sneu \ln\left(1-e^{-t\sqrt{\lambda_n}}\right), \label{p2}
\eeq
where we assume that $\lambda_n$ are the eigenvalues of a positive
definite elliptic differential operator $L$, acting on a
$p$-dimensional
Riemannian manifold $\cam$ with smooth boundary.
In order to motivate the following considerations, let us give
an example of how this quantity actually arise in concrete
field-theoretic problems. Most frequently, sums like the one in
eq.~(\ref{p2}) appear in finite temperature quantum field theory.
It was especially in this context, that also another method involving
the commutation of two (or more) series has been developed. There,
similar integral representations than the one in eq.~(2.3) have been
used at some point (see for example \cite{new}). However, making
systematic use of the Mellin-Barnes type integrals from the beginning of
the calculation leads to further simplication of the analysis.

For definiteness let us consider a free massive scalar field in an
ultrastatic curved spacetime ${\cal M}$ (possibly with boundary)
\cite{fulling,dowkerkennedy78}, to which we will apply the results
derived in this chapter.
The free energy of this system is defined by
\beq
F[\beta ] =-\frac 1 {\beta} \ln{\mbox T}{\mbox r} \exp [-\beta H],
\label{freeen} \eeq
with the Hamiltonian $H=\sum_j E_j [N_j +1/2]$ and the inverse
temperature $\beta$. Here $E_j$ are the energy eigenvalues determined by
$(-\Delta +\xi R +m^2) \psi _j =E_j^2 \psi _j$ with the Riemannian
curvature $R$ of ${\cal M}$, furthermore $-\Delta$ is the
Laplace-Beltrami
operator of the spatial section and $m$ is the mass of the field. The
part $(1/2)\sum_jE_j$ of the Hamiltonian $H$ is usually called the zero
point energy of the field. The operator $N_j$ is the number operator
associated with $\psi_j$ and the trace in (\ref{freeen}) has to be taken
over the Fock space of the field defined through the modes $\psi _j$.

A formal calculation then yields \cite{free}
\beq
F[\beta ] =\frac 1 2 \sum_j E_j +\frac 1 {\beta} \sum_j
 \ln\left(1-e^{-\beta E_j}\right),\label{formal}
\eeq
with the divergent zero-point energy and the finite temperature part
being exactly of the form (\ref{p2}). As is seen, the limit $t\to 0$ in
(\ref{p2}) corresponds to the high temperature limit in (\ref{formal}).

Another quantity associated with (\ref{p2}) is the generalized
generating function
\beq
Z(t)=\prod_n \left(1-e^{-t\sqrt{\lambda_n}}\right)^{-a},
\label{p1}
\eeq
where $a$ is a real number. The
knowledge of its asymptotic behaviour for small $t$ is
relevant,  for example, in obtaining
 the asymptotic state density behaviour of
$p$-branes.

Let us now consider $ G(t)$. The Mellin-transformation technique, once
more,
gives very easily the asymptotic expansion. First, by expanding the logarithm
and using (\ref{h2}), one has
\beq
 G(t)=-\frac 1 {2\pi i}\cont\zeta_R(1+\alpha )
t^{-\alpha}\zeta\left(L,\frac{\alpha} 2\right),
\label{p3}
\eeq
where
\beq
\zeta(L,\nu) =\sneu \lambda_n^{-\nu}\label{p4}
\eeq
is the zeta-function associated with the sequence of eigenvalues
$\{\lambda_n\}_{n\in\nats}$. In order to interchange $\sum_n$ and the
integral, one has to choose $c>p/2$.

For the evaluation of (\ref{p3}) one has to know the meromorphic
structure of $\zeta(L,\nu)$. General zeta-function theory (see for
example \cite{zeta}) tells
us that its poles of order one are located at $\nu =p/2,(p-1)/2,...,1/2;
-(2l+1)/2,l\in\nats_0$ (we will denote the corresponding
residues by
$R_{\nu}$ and the finite part by $C_{\nu}$). Thus one has the following
five posible types of poles enclosed on the left of the contour:
\begin{enumerate}
\item
$\alpha=p,p-1,...,1$: pole of order one due to $\zeta(L,\alpha /2)$.
\item
$\alpha=0$: pole of order two due to $\Gamma(\alpha)$ and
$\zeta_R(1+\alpha)$.
\item
$\alpha =-1$: pole of order two due to $\Gamma(\alpha)$ and $\zeta(L,\alpha
/2)$.
\item
$\alpha =-2k$, $k\in\nats$: pole of order one due to $\Gamma (\alpha )$.
\item
$\alpha =-(2k+1)$, $k\in\nats$: pole of order one due to the poles of
$\Gamma(\alpha)$ and $\zeta(L,\alpha/2)$ and the zero of $\zeta_R(1+\alpha)$.
\end{enumerate}
Summing over all contributions, one arrives at
\beq
 G(t) &\sim& -2\sum_{l=1}^p\Gamma(l)\zeta_R(1+l)t^{-l}R_{\frac l
2}\nn\\ & &-\left[\frac 1 2 \zeta'(L,0)-(\ln t)\zeta(L,0)\right]\label{p5}\\
& &-\left\{\frac 1 2 C_{-\frac 1 2}+R_{-\frac 1 2}\left[\ln(2\pi)+\psi
(2)-\ln t\right]\right\}t\nn\\
& &-\sleu \frac{t^{2l}}{(2l)!}\zeta_R(1-2l) \zeta(L,-l)\nn\\
& &+2\sleu \frac{t^{2l+1}}{(2l+1)! }\zeta_R'(-2l)R_{-\frac{2l+1} 2},\nn
\eeq
where the contour contributions which are exponentially damped for $t\to
0$ have not been written explicitly.

\section{First physical applications: a high-tempera\-ture expansion and
the asymptotic state density of $p$-branes}
\setcounter{equation}{0}
As already mentioned, a first physical application is quite immediate.
We can provide the explicit form of
the high-temperature expansion of a free (massive) scalar field in an
ultrastatic, curved
spacetime $\cam$ (possibly with boundary).
As briefly described, the free energy for this system is
\beq
F[\beta]=\frac 1 2 \sum_j E_j+F^{(\beta)}, \label{p6a}
\eeq
with
\beq
F^{(\beta)}=
\frac 1 {\beta}\sum_j \ln\left(1-e^{-\beta E_j}\right), \label{p6}
\eeq
where the small $t$-expansion in eq.~(\ref{p2}) here corresponds
to the high-tempera\-ture limit. By expressing the relevant information
of
the zeta-function $\zeta(L,\nu)$ in terms of the heat-kernel coefficients,
\beq
K(t)=\sum_j e^{-tE_j^2}\sim\left(\frac 1 {4\pi t}\right)^{\frac p 2}
\sum_{l=0,1/2,1,...}^{\infty}b_lt^l,\label{p7}
\eeq
that is
\beq
R_s=\frac{b_{\frac p 2 -s}}{(4\pi)^{\frac p 2}\Gamma(s)},\qquad
\zeta(L,-l)=(-1)^ll!\frac{b_{\frac p 2 +l}}{(4\pi)^{\frac
p2}},\label{p8} \eeq
the complete high-temperature expansion of the free energy may be found.
Using the doubling and the reflection formula for the $\Gamma$-function,
it reads
\beq
F^{(\beta)}&=&-\frac 1 2 PP\zeta(L,-1/2)+\frac 1 {(4\pi)^{\frac{p+1}
2}}\label{p9}\\
& &\times\left\{-b_{\frac{p+1}
2}\ln\left(\frac{\beta}{2\pi}\right)+\psi (2)+
\frac{2\sqrt{\pi}}{\beta}b_{\frac p 2}\ln\beta +P+S\right\},\nn
\eeq
with
\beq
S=-\sum_{r=1/2,1,...}^{\infty}b_{\frac{p+1} 2
+r}\left(\frac{\beta}{4\pi}\right)^{2r}\frac{(2r)!}{\Gamma(r+1)}\zeta_R
(1+2r)\label{p10}
\eeq
and
\beq
P=-\sum_{r=0,1/2,1,...}^{\frac{d-1}2}b_r\left(\frac{\beta}2\right)^
{-p-1+2r} \Gamma\left(\frac{p+1}2 -r\right)\zeta_R(p+1-2r),\label{p11}
\eeq
where $PP$ denotes the finite part of $\zeta(L,\nu)$.
This is certainly the result
previously found in \cite{dowkerkennedy78}.

Another application has to do with the calculation of the asymptotic
state density of $p$-branes.
Let us briefly consider it. The partition function associated with a
generating function of the kind $ G(t)$ may be written as
\beq
Z(z)=e^{-a G(z)}=\sum_n d_n e^{-nz}\, ,
\label{z}
\eeq
where $z$ is a complex variable, $z=t+iy$. The problem is to find
the asymptotic behavior of $d_n$ for large $n$. This can be accomplished
by making use of the asymptotic behaviour found in Eq. (\ref{p5}). The
Cauchy integral theorem gives
\beq
d_n=\frac{1}{2\pi i}\oint dz \, e^{nz}Z(z)\, ,
\eeq
where the countour integral consists of a small circle around the
origin. For $n$ very large, the leading contribution comes from the
asymptotic behaviour of $Z(z)$ for $z $ small. As a consequence, we
may write
\beq
d_n \simeq \frac{A}{2\pi i}\oint dz\, z^B e^{zn+Cz^{-p}}\, ,
\eeq
where
\beq
A=\exp\left\{\frac{a}{2}\zeta'(L,0)\right\}\, ,
\eeq
\beq
B=-a \zeta(L,0)\, ,
\eeq
\beq
C=2a \Gamma(p)\zeta(1+p)R_{\frac{p}{2}}\, .
\eeq
A straightforward calculation, based on a standard saddle point
technique, gives
\beq
d_n \simeq \frac{A}{\sqrt{2\pi (p+1)}}(pC)^{\frac{2B+1}{2(p+1)}}
n^{-\frac{2B+2+p}{2(p+1)}}
\exp\left\{\frac{p+1}{p}(pC)^{\frac{1}{p+1}}n^{\frac{p}{p+1}}\right\} \,
, \eeq
which generalizes Meinardus theorem \cite{meinardus54} (see also the
recent article by Actor \cite{mein}).
In the particular case of the semiclassical quantization of a $p$-brane,
compactified
on the torus $T^p$, this result leads to the asymptotic behaviour of
the corresponding level state density for large values of the mass (see
\cite{bytsenkokirstenzerbini93}).

\section{Application to the asymptotics of determinants in string
theory} \setcounter{equation}{0}
The last application concerns the asymptotics of the determinants which
appear in string theory. It is  well-known that the genus-$g$
contribution to the Polyakov bosonic string partition function
 can be written as \cite{dhokerphong88}
\beq
Z_{g} = \int (d\tau)_{WP}\, (\det P^+P)^{1/2}(\det \Delta_g)^{-13},
\label{1g}
\eeq
where $(d\tau)_{WP}$ is the Weil-Petersson measure on the
Teichm\"{u}ller
space, and $(\det P^+P)$ and $\det \Delta_g$ are the scalar and ghost
determinants, respectively. For our
purposes, here it is sufficient to observe that the integrand can be
expressed as \cite{dhokerphong88}
\beq
I_g(\tau)=(\det P^+P)^{1/2}(\det
\Delta_g)^{-13}=e^{c(2g-2)}Z'(1)^{-13}Z(2)\,, \label{2g}
\eeq
where $Z(s)$ is the Selberg zeta-function and $c$ is a constant.
We recall that the Selberg zeta-function, for untwisted scalar fields
(character $\chi (\gamma)=1$),
is defined by (for $\Re s>1$)
\beq
Z(s)=\prod_{\gamma}\prod_{n=0}^{\infty}(1-e^{-(s+n)l_{\gamma}})\,,
\label{3g}
\eeq
here the product is over primitive simple closed geodesics $\gamma$ on a
Riemann surface and $l_{\gamma}$ is the corresponding length.
Furthermore, $Z(s)$
is an entire function, non vanishing at $s=2$ but which has a simple
zero at $s=1$ \cite{hejhal76}, corresponding to
the zero mode of the scalar Laplace operator.

It is also known that physical quantities in string theory are expressed
as integrals over the moduli space. The integrands are regular
inside, thus, the only possible divergences have to be associated with
the  asymptotics near the boundary of the moduli
space. It is possible to show that this boundary corresponds to the
length of some geodesic
tending to $0$. So we are led to investigate the asymptotics for the
Selberg zeta-function and its derivatives. For the sake of simplicity,
let us consider one
pinching geodesic $\gamma_1$ and denote by $l_1$ its  length.
Since  $\gamma_1$  and its inverse are counted as distinct primitive
geodesics, we can write
\beq
Z(s)=R(s)\prod_{n=0}^{\infty}(1-e^{-(s+n)l_1})^2\,,
\label{3g3}
\eeq
with $R(s)$ bounded \cite{wolpert}, and the problem reduces to find the
asymptotic
behavior of the quantities $Z(2)$ and $Z'(1)$ when $l_1 $ goes to $0$.
With regard to the first quantity, this can be easily accomplished by
means of the technique
we have introduced. In fact, we obtain
\beq
\ln{\frac{Z(s)}{R(s)}}=-\frac{2}{2\pi i}\int_{\Re z>1} dz
\Gamma(z)\zeta(1+z) \zeta_H(z;s)l_1^{-z}.
\label{g4}
\eeq
The simple pole at $z=1$ and the double pole at $z=0$ give the leading
contributions, namely
\beq
\ln{\frac{Z(s)}{R(s)}}\simeq -\frac{2\zeta(2)}{l_1}+
\ln{l_1^{2\zeta_H(0;s)}}\,.
\label{g55}
\eeq
As a result
\beq
Z(s)\simeq R(s)e^{-\frac{\pi^2}{3l_1}}l_1^
{2\zeta_H(0;s)}\,.
\label{g5}
\eeq
With $\zeta_H(0;s)=1/2 -s$,
we have in particular
\beq
Z(2)\simeq R(2)e^{-\frac{\pi^2}{3l_1}}l_1^{-3}
\label{g6}
\eeq
and
\beq
Z(1+\epsilon)\simeq R(1+\epsilon)e^{-\frac{\pi^2}{3l_1}}l_1^{-1}\,.
\label{g7}
\eeq
Similar consideration are valid for the spinor sector, which
is relevant, e.g.~for super-strings. In this case, we have
for $\Re s>1$ (for details we refer the reader to
\cite{dhokerphong88}) \beq
Z_{1}(s)=\prod_{\gamma}\prod_{n=0}^{\infty}\left[1-\chi(\gamma)
e^{-(s+n)l_{\gamma}}\right]\,.
\label{3g2}
\eeq
Here the product is over primitive simple closed geodesics  $\gamma$ on
a Riemann surface, $l_{\gamma}$ is the corresponding length and
the character $\chi(\gamma)=\pm 1$ depends on the spin structure.
 Furthermore,
$Z_{1}(s)$
is non vanishing at $ s=3/2$, but has a  zero of order $2N$ at $s=1/2$, $N$
being the number of zero modes of the Dirac $D$ operator. We also have,
for the gravitino ghosts determinant
\beq
\det \left(P^{+}_{1/2}P_{1/2}\right)=e^{c_{1/2}}Z_{1}(3/2),
\eeq
and for the square of the Dirac determinant
\beq
\det D^2=e^{c_{D}}\frac{Z_{1}^{(2N)}(1/2)}{2N!}\,.
\eeq
Above, $c_{1/2}$ and $c_D$ are constants. Again the singularities are near the
boundary of the moduli
space and this boundary corresponds to the length of some geodesic
tending to $0$. Let $l_1$ be this length. The Ramond case $
(\chi(\gamma)=1$) is formally
similar to the untwisted scalar case. So let us consider the
antiperiodic case (Neveu-Schwarz), i.e.~$\chi(\gamma)=-1$.
We have
\beq
Z_{1}(s)=R_1(s)\prod_{n=0}^{\infty}(1+e^{-(s+n)l_1})^2\,,
\label{3gm}
\eeq
 with $R_1(s)$ bounded, and the problem reduces to find the asymptotic
behavior
of the quantities $Z_{1}(s)$ and of its derivatives at certain
points, when $l_1 $ goes to $0$.
Again, the Mellin technique is useful. It gives
\beq
\ln{\frac{Z_{1}(s)}{R_{1}(s)}}=\frac{1}{\pi i}\int_{\Re z>1} dz\,\,
\Gamma(z) (1-2^{-z})\zeta(1+z)
\zeta_H(z;s)l_1^{-z}
\label{g44}
\eeq
Due to the presence of the factor $ 1-2^{-z}$, we have simple poles at
$z=0,\, 1$. Thus, the leading
contributions are
\beq
\ln{\frac{Z_{1}(s)}{R_{1}(s)}}\simeq \frac{\zeta(2)}{l_1}
+2\zeta_H(0;s)\ln 2\,,
\label{g54}
\eeq
and as a result
\beq
Z_{1}(s)\simeq R_{1}(s)e^{\frac{\pi^2}{6l_1}} 2^
{2\zeta_H(0;s)}\,.
\label{g555}
\eeq
In particular, we have
\beq
Z_{1}(3/2)\simeq \frac 1 4 R_{1}(3/2)e^{\frac{\pi^2}{6l_1}}.
\label{g66}
\eeq
The asymptotic behaviour of the derivative at $s=1$ and  $s=1/2$ are
more difficult to investigate. A possible approach consists in
starting from \cite{hejhal76}
\beq
\frac{Z'(s)}{Z(s)}=\sum_{\gamma}\sum_{n=1}^\infty
\frac{l_{\gamma}}{2\sinh {\frac{nl_{\gamma}}{2}}}e^{-(s-1/2)n\l_{\gamma}}\,.
\label{dz}
\eeq
Separating here the contribution of the shrinking geodesic, we have
\beq
\frac{Z'(s)}{Z(s)}=
\sum_{n=1}^\infty
\frac{l_1}{\sinh {\frac{nl_{1}}{2}}}e^{-(s-1/2)n\l_1}+H(s)\,.
\label{dz1}
\eeq
Again, the Mellin technique allows us to write
\beq
\frac{Z'(s)}{Z(s)}=\frac{l_1}{\pi i}\int_{\Re z>2} dz \Gamma(z)
\zeta(z) \zeta_H(z;s)l_1^{-z}+H(s)\, .
\label{g444}
\eeq
Note that one can arrive at the same result, taking the derivative of
Eq.~(\ref{g44}) with respect to $s$ and making use of
$z\Gamma(z)=\Gamma(z+1)$ and \beq
\frac{\abl}{\abl s}\zeta_H(z;a)&=&-z\zeta_H(z+1;s).
\eeq
Thus the analogue result for the Neveu-Schwarz spinor is simply
\beq
\frac{Z_1'(s)}{Z_1(s)}=\frac{-l_1}{\pi i}\int_{\Re z>2} dz \Gamma(z)
\zeta(z) \zeta_H(z;s)l_1^{-z} (1-2^{1-z})+H_1(s)\, .
\label{g4445}
\eeq

In the first case, the integrand has a double pole at $z=1$ and simple poles at
$z=0,-1,-2,..$. Thus, the leading contribution is
\beq
\frac{Z'(s)}{Z(s)}=-2[\ln l_1+\psi(s)]+H(s)+O(l_1)\, .
\label{g4444}
\eeq
In the second case, we have simple poles at $z=1, 0, -1,...$ and the
leading term is
\beq
\frac{Z_1'(s)}{Z_1(s)}=-2\ln 2+H_1(s)+O(l_1)\, .
\label{g44446}
\eeq
Let us briefly discuss the bosonic case. For $\Re s>1$, there are no
problems and the above result gives us the
asymptotic behavior for the logarithmic derivative of the Selberg
zeta-function,
when $l_1$ goes to $0$. The delicate point is $s=1$, where $Z(s)$
has a simple zero.
However, by using the analytical properties of the Selberg
zeta-function, one may first show,
$ R(1+\epsilon)H(1+\epsilon)\simeq  B(1)+O(\epsilon)$, and thus we get
\beq
Z'(1)\simeq B(1)e^{-\frac{\pi^2}{3l_1}}l_1^{-1}
\label{g8}
\eeq
which gives the correct leading contribution \cite{wolpert}.
We conclude by saying that the asymptotic behaviors found for $Z(2)$ and
$Z'(1)$ lead to the celebrated double-pole theorem of Belavin
and Knizhnik \cite{belavinknizhnik86} for the quantum bosonic string.

\section{Conclusions}
In this article we have applied the Mellin-Barnes integral
transformation
to several situations of present interest in mathematical physics.
We have shown, in our opinion, that this technique is best suited
for the derivation of asymptotic expansions in different contexts.
Compared with other approaches, it is certainly more simple and
straightforward. Moreover, the
expansions we have obtained may be relevant in their own right as we
have seen in sections
4 to 6. Supplemented by numerical techniques and explicit results that
have become handy recently---adapted to the notion of
asymptotic approximation---our method also constitutes a useful device
in
order to provide actual numbers to be contrasted with experimental
measurements in
fields like the determination of the Casimir energy or the study of
actual physical implications of some Kaluza-Klein theories
\cite{4,5}.

\section*{Acknowledgments}
KK would like to thank the members of the Department ECM,
Barcelona University, for the kind hospitality.
Furthermore, we thank the referees for valuable comments.
This work has been supported by the Alexander von Humboldt
Foundation
(Germany), by DGICYT (Spain), project PB93-0035, and by CIRIT
(Generalitat de Catalunya).

\end{document}